\makeatletter
\newcommand{\dontusepackage}[2][]{%
  \@namedef{ver@#2.sty}{9999/12/31}%
  \@namedef{opt@#2.sty}{#1}}
\makeatother
\dontusepackage{subfigure}

\documentclass[]{article}

\usepackage{lmodern}
\usepackage{amssymb,amsmath}
\usepackage{ifxetex,ifluatex}
\usepackage[usenames,dvipsnames]{color}
\usepackage{fixltx2e} % provides \textsubscript
\ifnum 0\ifxetex 1\fi\ifluatex 1\fi=0 % if pdftex
  \usepackage[T1]{fontenc}
  \usepackage[utf8]{inputenc}
\else % if luatex or xelatex
  \ifxetex
    \usepackage{mathspec}
    \usepackage{xltxtra,xunicode}
  \else
    \usepackage{fontspec}
  \fi
  \defaultfontfeatures{Mapping=tex-text,Scale=MatchLowercase}
  
\fi
\IfFileExists{upquote.sty}{\usepackage{upquote}}{}
\IfFileExists{microtype.sty}{%
\usepackage{microtype}
\UseMicrotypeSet[protrusion]{basicmath} % disable protrusion for tt fonts
}{}
\usepackage[margin=1.0in,bottom=1.5in]{geometry}
\usepackage[square,numbers,sort&compress]{natbib}
\usepackage{listings}
\usepackage{graphicx}
\makeatletter
\def\maxwidth{\ifdim\Gin@nat@width>\linewidth\linewidth\else\Gin@nat@width\fi}
\def\maxheight{\ifdim\Gin@nat@height>\textheight\textheight\else\Gin@nat@height\fi}
\makeatother
\setkeys{Gin}{width=\maxwidth,height=\maxheight,keepaspectratio}
\usepackage{caption}
\usepackage{float}
\usepackage{subfig}
\ifxetex
  \usepackage[setpagesize=false, % page size defined by xetex
              unicode=false, % unicode breaks when used with xetex
              xetex]{hyperref}
\else
  \usepackage[unicode=true]{hyperref}
\fi
\hypersetup{breaklinks=true,
            bookmarks=true,
            pdfauthor={},
            pdftitle={Wave-equation-based inversion with amortized variational Bayesian inference},
            colorlinks=true,
            citecolor=black,
            urlcolor=blue,
            linkcolor=black,
            pdfborder={0 0 0}}
\usepackage[preprint]{neurips_2019}
\usepackage{url}
\usepackage{booktabs}
\usepackage{amsfonts}
\usepackage{nicefrac}
\usepackage{microtype}

\newcommand{\R}{\mathrm}
\newcommand{\B}{\mathbf}

\DeclareMathOperator*{\argmin}{arg\,min}

\title{Wave-equation-based inversion with amortized variational Bayesian
inference}
\author{Ali Siahkoohi\\School of Computational Science and Engineering,\\Georgia
Institute of Technology\\\texttt{alisk@gatech.edu}\\\And
Rafael Orozco\\School of Computational Science and Engineering,\\Georgia
Institute of Technology\\\texttt{rorozco@gatech.edu}\\\And
Gabrio Rizzuti\\Department of Mathematics,\\Utrecht
University\\\texttt{g.rizzuti@uu.nl}\\\And
Felix J. Herrmann\\School of Computational Science and
Engineering,\\Georgia Institute of
Technology\\\texttt{felix.herrmann@gatech.edu}}
\date{}

\begin{document}
\maketitle
\begin{abstract}
Solving inverse problems involving measurement noise and modeling errors
requires regularization in order to avoid data overfit. Geophysical
inverse problems, in which the Earth's highly heterogeneous structure is
unknown, present a challenge in encoding prior knowledge through
analytical expressions. Our main contribution is a
generative-model-based regularization approach, robust to
out-of-distribution data, which exploits the prior knowledge embedded in
existing data and model pairs. Utilizing an amortized variational
inference objective, a conditional normalizing flow (NF) is pretrained
on pairs of low- and high-fidelity migrated images in order to achieve a
low-fidelity approximation to the seismic imaging posterior distribution
for previously unseen data. The NF is used after pretraining to
reparameterize the unknown seismic image in an inversion scheme
involving physics-guided data misfit and a Gaussian prior on the NF
latent variable. Solving this optimization problem with respect to the
latent variable enables us to leverage the benefits of data-driven
conditional priors whilst being informed by physics and data. The
numerical experiments demonstrate that the proposed inversion scheme
produces seismic images with limited artifacts when dealing with noisy
and out-of-distribution data.
\end{abstract}

\section{Introduction}\label{introduction}

An inverse problem involves reliably estimating an unknown quantity from
noisy indirect observations. This problem is commonly solved using
optimization techniques to minimize the difference between predicted and
observed data. Solely minimizing the data misfit negatively impacts the
quality of the obtained solution due to noise in the data, modeling
errors, and a nontrivial null-space of the forward operator
\citep{tarantola2005inverse}. To prevent this, it is crucial to capture
and incorporate prior knowledge into the inverse problem
\citep{tarantola2005inverse}, e.g., Gaussian or Laplace distribution
priors
\citep{malinverno2004expanded, MartinMcMC2012, herrmann11GPelsqIm}.
While theoretically understood, these type of priors may lead to
undesirable biases in the outcome of inversion.

The purpose of our contribution is to address this challenge by
utilizing a formulation that exploits a data-driven conditional prior.
To achieve this, following \citet{orozco2021photoacoustic}, we train a
conditional normalizing flow \citep[NF,][]{kruse2021hint} to capture the
conditional distribution of the unknown, given data, i.e., the posterior
distribution. The training involves minimizing an amortized variational
inference objective
\citep{jordan1999introduction, siahkoohi2020ABIpto, siahkoohi2021Seglbe, kruse2021hint, kovachki2021conditional}
using existing training pairs in the form of low-fidelity data and model
pairs. After training, we are able to capture the low-fidelity posterior
distribution for previously unseen seismic data. We use the this network
to reparameterize the unknown in an inversion scheme, involving
physics-guided data misfit and a Gaussian prior on the NF latent
variable. Due to the inherent invertibility of NFs, they can represent
any model in the unknown space, which allows them to be be used as
priors when dealing with out-of-distribution data
\citep{pmlr-v119-asim20a, orozco2021photoacoustic}.

There are three key advantages to our proposed method: (1) Data-driven
priors make use of available data, such as high-resolution seismic
images to capture prior knowledge about the Earth's subsurface; (2) The
use of a conditional prior favors solutions that are consistent with the
data, which provides more specific knowledge about the unknown; (3) With
the help of our formulation, we combine data-driven priors with
conventional physics-based inversion methods, which offers the
advantages of data-driven priors without relying solely on them as a
black box.

In the following sections, we discuss conditional NFs, trained using an
amortized variational inference procedure. Next, we present an inversion
scheme for seismic imaging that incorporates conditional priors. We
conclude by demonstrating this technique on a realistic seismic imaging
problem involving noisy and out-of-distribution data.

\section{Amortized variational
inference}\label{amortized-variational-inference}

The problem setup entails applying variational inference
\citep{jordan1999introduction} to approximate the posterior distribution
\citep{tarantola2005inverse} associated with the inverse problem
$\B{y} = F (\B{x}) + \boldsymbol{\epsilon}$, where $\B{y} \in Y$
represents the observed data, $\B{x} \in X$ unknown model,
$\boldsymbol{\epsilon}$ possibly non-Gaussian noise, and
$F: X \rightarrow Y$ the possibly nonlinear forward operator. In the
context of amortized variational inference \citep{kruse2021hint}, we
wish to approximate the posterior distribution associated with this
inverse problem for previously unseen data. This method has
computational advantages as it does not require solving an additional
instance of variational inference for new data. In this work, we choose
NFs \citep{rezende2015variational} that due to their invertibility (up
to numerical precision) can be used to approximate a target
distribution, of which we have only samples. NFs can be adapted to
sample from the conditional distribution $p(\B{x} \mid \B{y})$ by using
a block-triangular construction \citep{marzouk2016sampling},
$T_{\R{w}} (\B{y}, \B{x}) =(T_{\R{w}_1} (\B{y}), T_{\R{w}_2} (\B{y}, \B{x}))$
with $\B{w} = (\B{w}_1, \B{w}_2)$. The conditional NF
$T_{\R{w}} : Y \times X \rightarrow Z \times Z$, which takes as input
data and model pairs $(\B{y}, \B{x})$, aims to yield two normally
distributed outputs in the latent space $Z \times Z$. Training objective
is based on minimizing the Kullback-Leibler divergence between the NF
output distribution and the Gaussian latent distribution
\citep{kruse2021hint}:
\begin{equation}
\mathop{\rm arg\,min}_{\B{w}}\, \frac{1}{n}
    \sum_{i=1}^n \Bigg[ \frac{1}{2}\big\|
    T_{\R{w}} (\B{y}^{(i)}, \B{x}^{(i)}) \big\|^2_2 -
    \log \Big | \det \nabla_{(\R{y}, \R{x})} T_{\R{w}} (\B{y}^{(i)},
    \B{x}^{(i)}) \Big |\Bigg].
\label{NF-training-cond}
\end{equation}
 In the above objective, the $\ell_2$-norm follows from a Gaussian
assumption on the latent variables and the second term is a
regularization term that avoids $T_{\R{w}}$ from converging to trivial
solutions---e.g., $T_{\R{w}} := \B{0}$. Computing
$\det \nabla_{(\R{y}, \R{x})} T_{\R{w}} (\B{y}, \B{x})$ and its gradient
adds almost no extra cost because of the particular design of invertible
networks \citep{dinh2016density}. Following training, we can obtain
samples from conditional distribution $p(\B{x} \mid \B{y})$ via
$T_{\R{w}_2}^{-1} \big( T_{\R{w}_1} (\B{y}), \B{z} \big) \sim p(\B{x} \mid \B{y}), \ \B{z} \sim p_{\R{z}}(\B{z})$
\citep{marzouk2016sampling, kruse2021hint}. This amounts to feeding the
latent code associated with observed data, i.e., $T_{\R{w}_1} (\B{y})$,
and Gaussian samples $\B{z} \sim p_{\R{z}}(\B{z})$ into the inverse
network, $T_{\R{w}_2}^{-1}$. These samples may be used for Bayesian
inference if we have an ideal training dataset
\citep{herrmann2019NIPSliwcuc, siahkoohi2021Seglbe, kumar2021enabling, kothari2021trumpets}.
However, such an assumption is rarely correct in geophysical
applications due to Earth's strong heterogeneity
\citep{siahkoohi2019transfer, sun2020elastic, siahkoohi2021deep}, which
highlights the importance of devising formulations that are robust to
changes in data distribution during inference.

\section{Seismic imaging with data-driven conditional
priors}\label{seismic-imaging-with-data-driven-conditional-priors}

Using multiple processed shot records,
$\left \{\B{d}_{i}\right \}_{i=1}^{n_s}$, seismic imaging aims to
estimate the short-wavelength component of the Earth's subsurface
squared-slowness model, denoted by $\delta \B{m}$. The linearized Born
scattering operator $\B{J}(\B{m}_0, \B{q}_i)$ relates the unknown
seismic image $\delta \B{m}^{\ast}$, to data, the $i\text{th}$ source
signature, $\B{q}_{i}$, and the background squared-slowness model
$\B{m}_0$. This relation can be written as
\begin{equation}
\B{d}_i = \B{J}(\B{m}_0, \B{q}_i)
    \delta \B{m}^{\ast} + \boldsymbol{\epsilon}_i, \quad
    \boldsymbol{\epsilon}_i \sim \mathrm{N} (\B{0}, \sigma^2
    \B{I}).
\label{linear-fwd-op}
\end{equation}
 Noise is denoted by $\boldsymbol{\epsilon}_i$ and represents
measurement noise and linearization errors, which for simplicity is
assumed to be distributed according to a zero-centered Gaussian
distribution with known covariance $\sigma^2 \B{I}$. We train a NF on
pairs of low- and high-fidelity seismic images via the amortized
variational inference objective by choosing $\B{x}$ to represent
high-fidelity migrated images and $\B{y} := \delta\B{m}_{\text{RTM}}$
corresponding to low-fidelity reverse-time migrated images obtained by
the process of demigration, followed by adding noise and migration.
After training, the conditional NF captures the low-fidelity seismic
imaging posterior distribution. In order to obtain a high-fidelity
seismic image maximum a posteriori (MAP) estimate, we propose to
reparameterize $\delta \B{m}$ vis the pretrained NF and solve the
optimization problem
\begin{equation}
\widehat{\B{z}} = \argmin_{\B{z}}\,  \frac{1}{2 \sigma^2}
    \Bigg[ \sum_{i=1}^{n_s}\big \|
    \B{d}_i- \B{J}(\B{m}_0, \B{q}_i)
    T_{\R{w}_2}^{-1} \big( T_{\R{w}_1} (\delta\B{m}_{\text{RTM}}), \B{z} \big) \big \|_2^2 \Bigg] + \frac{1}{2} \big \|
    \B{z} \big \|_2^2,
\label{imaging-opt-nf}
\end{equation}
 followed by mapping
$\delta\B{m}_{\text{MAP}} := T_{\R{w}_2}^{-1} \big( T_{\R{w}_1} (\delta\B{m}_{\text{RTM}}), \widehat{\B{z}} \big)$.
We initialize optimization problem~\ref{imaging-opt-nf} with
$\B{z}_0 = \B{0}$. This initialization and a Gaussian prior on $\B{z}$
regularize the inversion by favoring solutions that are likely samples
of the low-fidelity posterior distribution \citep{pmlr-v119-asim20a}.
NFs' inherent invertibility allows them to represent any image
$\delta \B{m}$ in the solution space. This limits the potentially
negative bias of the conditional prior in domains where access to
high-fidelity training data is limited. We demonstrate this through a
numerical experiment in the next section.

\section{Numerical experiments}\label{numerical-experiments}

\begin{figure*}
\centering
\subfloat[\label{true-2}]{\includegraphics[width=0.450\hsize]{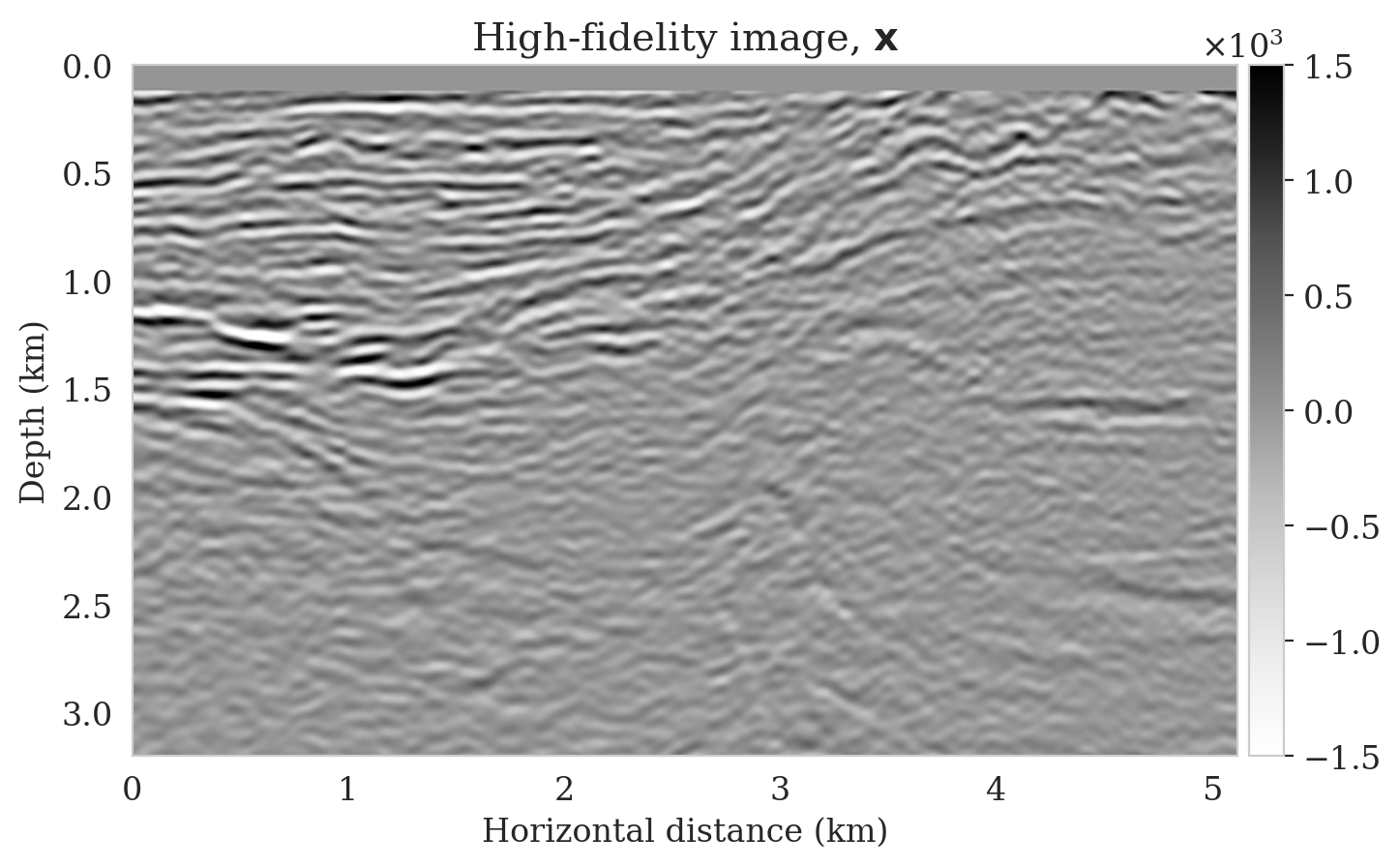}}
\subfloat[\label{rtm-2}]{\includegraphics[width=0.450\hsize]{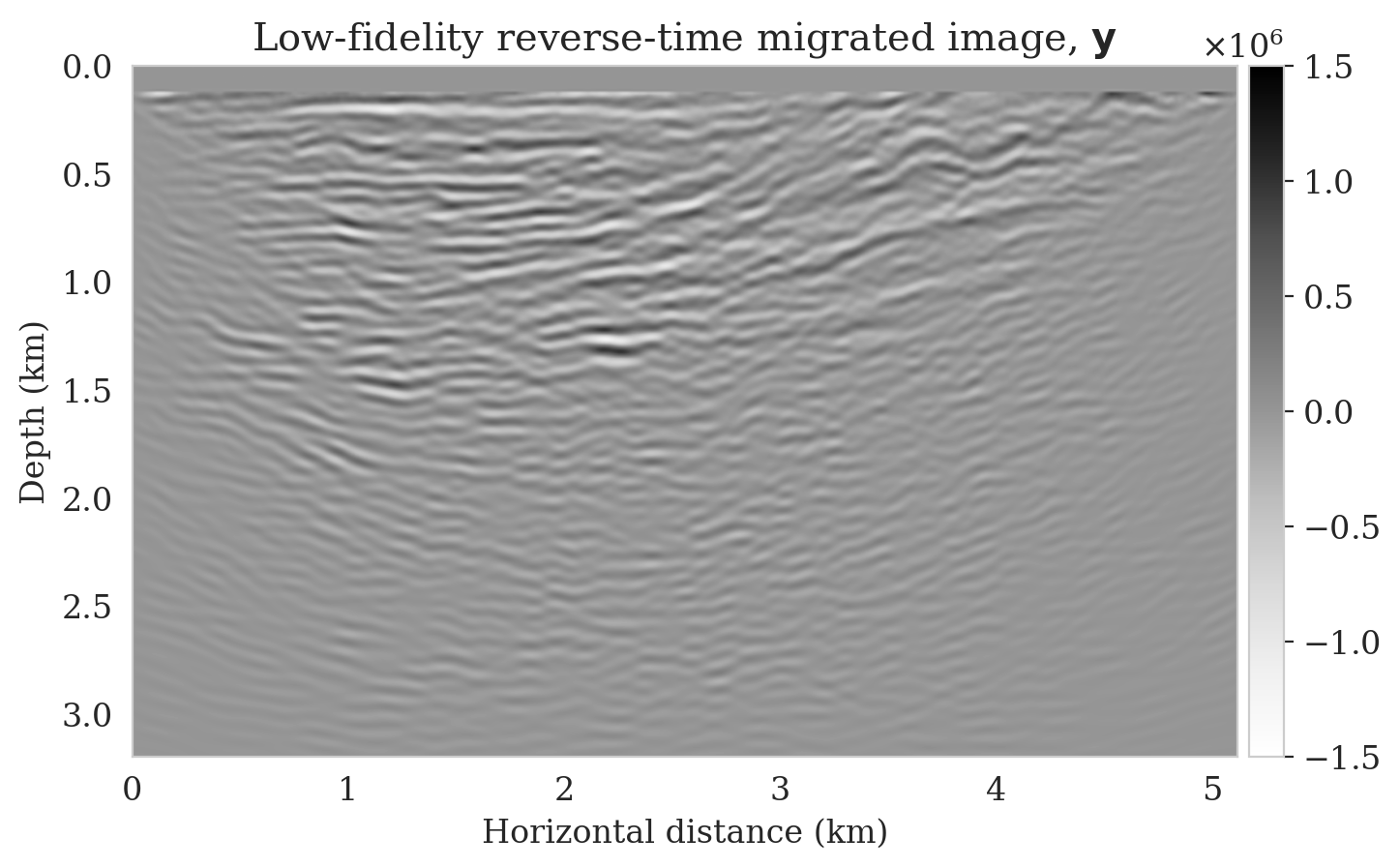}}
\\
\subfloat[\label{sample-2-4}]{\includegraphics[width=0.450\hsize]{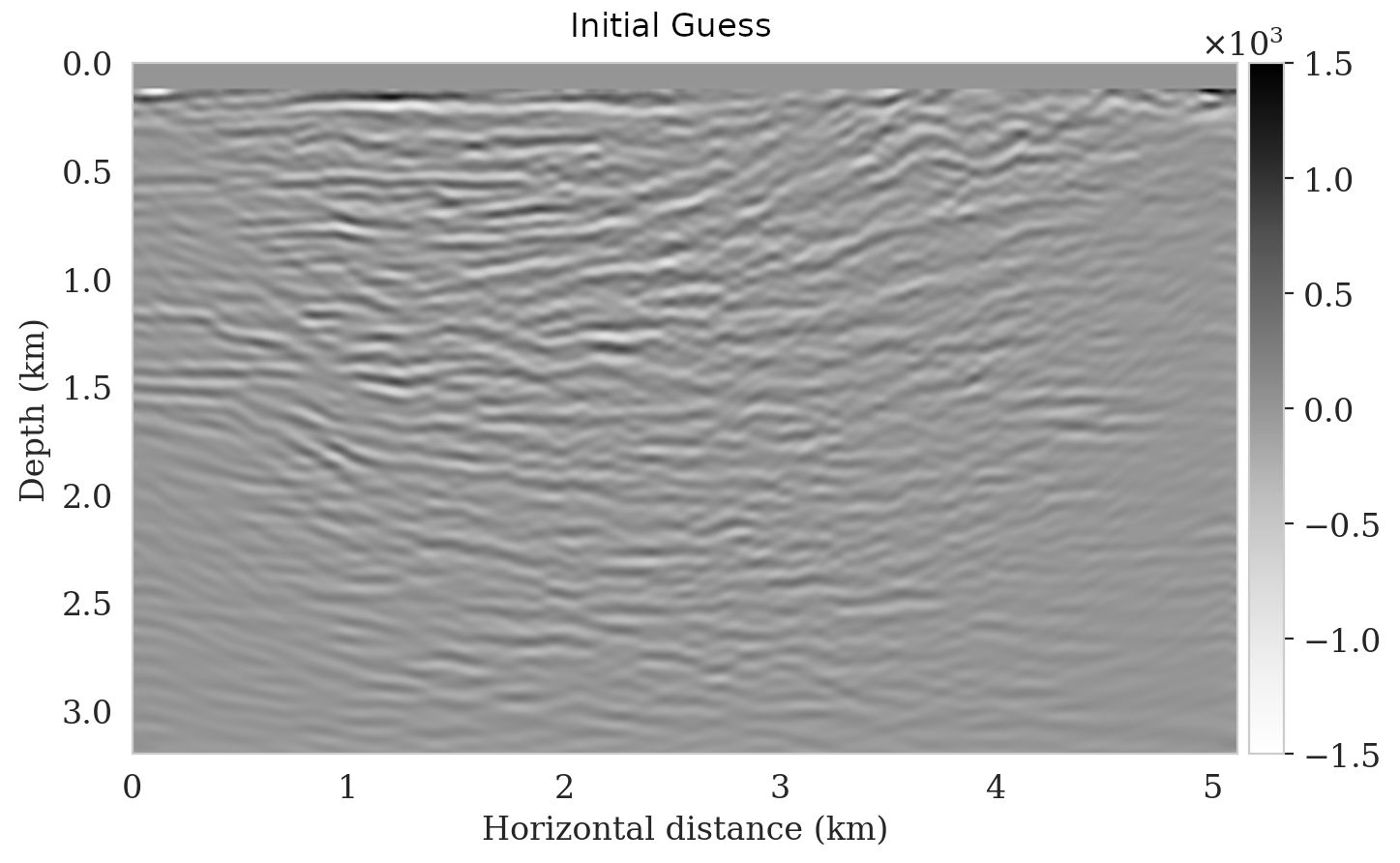}}
\subfloat[\label{cm-2}]{\includegraphics[width=0.450\hsize]{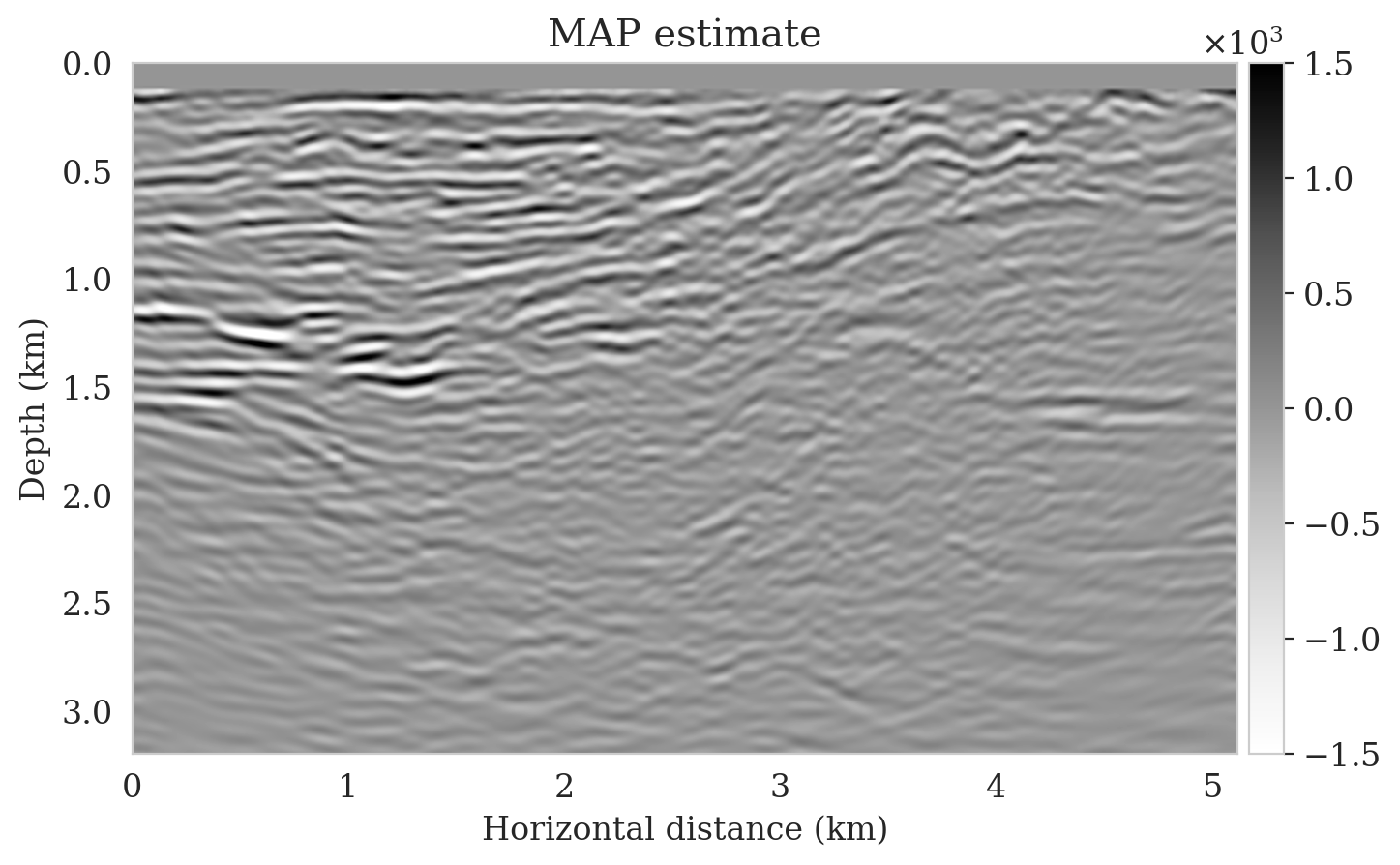}}
\caption{Imaging with conditional NF priors. (a) High-fidelity image.
(b) Reverse-time migrated image. (c) Initial guess,
$T_{\R{w}_2}^{-1} \big( T_{\R{w}_1} (\delta\B{m}_{\text{RTM}}), \B{z}_0 \big)$.
(d) MAP estimation with conditional NF prior
(Equation~\ref{imaging-opt-nf}).}\label{example}
\end{figure*}

We propose a realistic example in which we create $4750$ 2D training
pairs of low- and high-fidelity seismic images, which the latter are
$3075\, \mathrm{m} \times 5120\, \mathrm{m}$ sections extracted from the
shallow part of \href{https://wiki.seg.org/wiki/Parihaka-3D}{Parihaka}
\citep{WesternGeco2012} prestrack Kirchhoff migration dataset. The
low-fidelity images are obtain by migrating noisy synthetic data
obtained from the high-fidelity images according to
Equation~\ref{linear-fwd-op}. The acquisition geometry involves $102$
shot records, $204$ fixed receivers, Ricker wavelet with a central
frequency of $30\, \mathrm{Hz}$, and band-limited noise. We augment a
$125\, \mathrm{m}$ water column on top of these models to limit the near
source imaging artifacts. We train $T_{\R{w}}$ according to the
objective function in Equation~\ref{NF-training-cond} with the Adam
optimization algorithm \citep{kingma2014adam}. In order to evaluate the
effectiveness of our inversion scheme when applied to
out-of-distribution data, we select a 2D section from the deeper
portions of the Parihaka dataset (see Figure~\ref{true-2}). As compared
to training images, this image includes more noncontinuous reflectors,
due to low signal-to-noise ratio in the deeper parts of the Parihaka
dataset. We simulated linearized data with the same acquisition geometry
described above to obtain a low-fidelity image (Figure~\ref{rtm-2}). We
solve optimization problem~\ref{imaging-opt-nf} for $5$ passes over the
shot records, i.e., approximately the same cost as 5 reverse-time
migrations. Figure~\ref{sample-2-4} shows the initial guess of the
optimization, i.e.,
$T_{\R{w}_2}^{-1} \big( T_{\R{w}_1} (\delta\B{m}_{\text{RTM}}), \B{z}_0 \big)$.
We can see that this image is not correctly recovering the reflectors,
however, it can be considered a better starting guess than the
reverse-time migrated image (Figure~\ref{rtm-2}) due to corrected
amplitudes. Finally, Figure~\ref{cm-2} shows the MAP estimate, obtained
via solving the optimization problem~\ref{imaging-opt-nf}, which
successfully reconstructs most of the reflectors with limited artifacts.

Our example uses \href{https://github.com/slimgroup/JUDI.jl}{JUDI}
\citep{witteJUDI2019} to construct wave-equation solvers, which utilizes
\href{https://www.devitoproject.org/}{Devito}
\citep{devito-compiler, devito-api} as a highly optimized finite
difference solver under the hood. The network architectures are
implemented using
\href{https://github.com/slimgroup/InvertibleNetworks.jl}{InvertibleNetworks.jl}
\citep{invnet}, a memory-efficient framework for training invertible
nets in Julia. Sample code to reproduce the results is provided on
\href{https://github.com/slimgroup/ConditionalNFs4Imaging.jl}{GitHub}.

\section{Conclusions}\label{conclusions}

Considering the Earth's strong heterogeneity, designing regularization
schemes to incorporate prior knowledge for solving ill-posed geophysical
inverse problems is challenging. To address this challenge, we proposed
a regularization scheme that takes advantage of existing data in the
form of low- and high-fidelity seismic images to train a conditional
normalizing flow (NF). This conditional NF approximates the imaging
posterior distribution for previously unseen data. In order to minimize
the impact of data distribution shifts during inference, we
reparameterized the unknown image with the conditional NF and inverted
for a Gaussian latent variable that fits the data. The resulting maximum
a posteriori estimate takes advantage of the data-driven conditional
prior while remaining bound to data and physics. Using numerical
experiments, we demonstrated that this approach yields seismic images
with limited imaging artifacts in the absence of high-fidelity training
data. Further research on quantifying the uncertainty through this
regularization technique is required.

\section{Acknowledgment}\label{acknowledgment}

This research was carried out with the support of Georgia Research
Alliance and partners of the ML4Seismic Center.

\bibliography{abstract}

\end{document}